\documentclass[11pt]{article}

	\newcommand{\blind}{0}
	
    \makeatletter
    \renewcommand\section{\@startsection {section}{1}{\z@}%
                                       {-3.5ex \@plus -1ex \@minus -.2ex}%
                                       {2.3ex \@plus.2ex}%
                                       {\normalfont\fontfamily{phv}\fontsize{16}{19}\bfseries}}
    \renewcommand\subsection{\@startsection{subsection}{2}{\z@}%
                                         {-3.25ex\@plus -1ex \@minus -.2ex}%
                                         {1.5ex \@plus .2ex}%
                                         {\normalfont\fontfamily{phv}\fontsize{14}{17}\bfseries}}
    \renewcommand\subsubsection{\@startsection{subsubsection}{3}{\z@}%
                                        {-3.25ex\@plus -1ex \@minus -.2ex}%
                                         {1.5ex \@plus .2ex}%
                                         {\normalfont\normalsize\fontfamily{phv}\fontsize{14}{17}\selectfont}}
    \makeatother
	
	\usepackage{amsmath}
	\usepackage{graphicx}
	\usepackage{enumerate}
	\usepackage{xcolor}
    
    \usepackage{fancyhdr}                           
    \usepackage{graphicx}                               
	\usepackage{natbib} 
	\usepackage{url} 
	\newcommand{\bit}{\begin{itemize}}
    \newcommand{\eit}{\end{itemize}}
 
	 \usepackage{comment}    
    \usepackage{pdflscape}
    \usepackage{makecell}
    \usepackage{setspace}

\begin{document}

		\def\spacingset#1{\renewcommand{\baselinestretch}%
			{#1}\small\normalsize} \spacingset{1}

		\if0\blind
		{
			\title{\bf A Gateway to Quantum Computing \\for Industrial Engineering}
			\author{Emily L. Tucker $^a$ and Mohammadhossein Mohammadisiahroudi $^b$ \\
			$^a$ Department of Industrial Engineering, Clemson University, Clemson, SC, USA \\
             $^b$ Department of Mathematics and Statistics,\\ University of Maryland, Baltimore County, MD, USA }
			\date{}
			\maketitle
		} \fi

\begin{abstract}
Quantum computing is rapidly emerging as a new computing paradigm with the potential to improve decision-making, optimization, and simulation across industries. For industrial engineering (IE) and operations research (OR), this shift introduces both unprecedented opportunities and substantial challenges. The learning curve is high, and to help researchers navigate the emerging field of quantum operations research, we provide a road map of the current field of quantum operations research. We introduce the foundational principles of quantum computing, outline the current hardware and software landscape, and survey major algorithmic advances relevant to IE/OR, including quantum approaches to linear algebra, optimization, machine learning, and stochastic simulation. We then highlight applied research directions, including the importance of problem domains for driving long-term value of quantum computers and how existing classical OR models can be reformulated for quantum hardware. 
Recognizing the steep learning curve, we propose pathways for IE/OR researchers to develop technical fluency and engage in this interdisciplinary domain. By bridging theory with application, and emphasizing the interplay between hardware and research development, we argue that industrial engineers are uniquely positioned to shape the trajectory of quantum computing for practical problem-solving. Ultimately, we aim to lower the barrier to entry into QOR, motivate new collaborations, and chart future directions where quantum technologies may deliver tangible impact for industry and academia.

	\end{abstract}
			
	\noindent%

	\spacingset{1.5}

\section{Introduction} \label{s:intro}
A fundamentally new style of computing has entered the scene: quantum computers. They have received widespread attention, and governments worldwide are pouring billions of dollars into quantum hardware and software development \citep{erixon2025benchmarking}. There is an on-going race towards “quantum advantage” where a quantum-based system may outperform a classical system alone \citep{huang2025vast}. But what are quantum computers? And what do they mean for industrial engineering research? 

Quantum computers work by manipulating individual particles such as electrons, ions, or photons \citep{de2021materials}. 
While items of every size are affected by quantum mechanics, its probabilistic effects are  seen most prominently in the behavior of individual particles. A key insight in the development of quantum computers was that we can use principles of quantum mechanics as a feature to harness rather than an bug to design around. As such the calculations on a quantum computer are inherently probabilistic \citep{nielsen2010quantum}, which is very different from the deterministic processes on a classical computer. 
Three ideas from quantum physics will be particularly relevant in our discussion: entanglement, superposition, interference; we will discuss each in Section \ref{s:QCBasics}.

These ideas open new possibilities for how we can design algorithms and for how we may one day model systems. There is some optimism that hybrid quantum-classical systems may expedite solutions to problems that seek to improve decisions, predictions, and simulations. Quantum computers also induce new types of research questions for our discipline. These include direct questions such as how to leverage quantum mechanics principles for algorithm design and how to model problems such that they can be represented on quantum computers, as well as broader questions regarding what styles of existing problems have quantum potential, what is best kept on classical machines, and how to coordinate an industry where hardware and software are both rapidly and interdependently developing. Industrial engineers and operations researchers (IE/OR researchers) could play an important role in supporting the future of quantum computers for decision-making. 

The field of IE/OR has already made inroads in improving quantum computation. We can run discrete optimization problems on quantum computers, and we can predict \citep{biamonte2017quantum}. Research in quantum optimization \citep{abbas2024challenges} and quantum machine learning \citep{biamonte2017quantum} is active and on-going. The broader quantum community has recognized optimization as a major area in which quantum computers could lead to real-world benefit, and some companies have even started proof-of-concept implementations. Quantum annealing, one style of hybrid quantum-classical algorithm (Section \ref{s:QCOR}), has been used to manage shipping containers at a major port in Los Angeles \citep{dwave_port_la_pier300} and to schedule TV commercials on a Japanese network \citep{dwave_recruit_tvads}.

Beyond this general motivation, IE/OR holds a unique position in shaping the trajectory of quantum computing. As many investors and funding agencies prioritize demonstrable practical impact, IE/OR can play a pivotal role in showcasing how quantum computing can solve real-world problems, thereby accelerating the co-development of better quantum hardware and software \citep{bernal2025what}. Quantum computing also invites a re-examination of long-established IE/OR methodologies. Many of the modeling, algorithmic, and implementation paradigms that have guided decision science for decades may be reformulated, redesigned, or rethought to align with quantum architectures. 

For example, in classical optimization, linear programming has long served as the backbone of many algorithms.  
In contrast, several quantum optimization paradigms operate on quadratic unconstrained binary optimization (QUBO) formulations. This distinction introduces new research opportunities: how to efficiently reformulate practical constrained problems into QUBO form, how to exploit quantum QUBO solvers as algorithmic subroutines, and how to bridge linear and quadratic formulations in hybrid pipelines.

Moreover, IE/OR expertise can contribute back to the development of more effective quantum algorithms and architectures for practical applications. For instance, while noise and randomness in current quantum hardware are often viewed as challenges, recent studies have shown that such stochasticity can be harnessed as a feature in certain optimization or sampling contexts \citep{domino2025baltimore}. Co-designing quantum hardware with application-driven insight can yield mutual benefit, e.g., QuEra's  device architectures were tailored to solve combinatorial problems like the maximum independent set \citep{ebadi2022quantum}. Quantum-inspired algorithmic concepts are already informing the design of advanced classical methods; quantum-inspired GPU-enhanced prediction, for example, has been used for solar energy prediction \citep{hong2024solar}. 

In general, quantum IE/OR is at a nascent stage. Problems with 20 variables or fewer can still be challenging to solve, 
and many algorithms are not yet viable with existing hardware. 
Research and industrial work are needed to map the future of what quantum IE/OR could be. 
For IE/OR to be a major player in this space, we need to extend our reach as a discipline.  

However, there is a high barrier to entry into quantum IE/OR, which we will term QOR moving forward. 
While deep knowledge of quantum mechanics is not necessary, because quantum computing has been largely driven by quantum physicists, the working notation and conventions can be unfamiliar to those with an IE/OR background. 
Further, we have observed that much of the introductory literature in this space has either ramped too quickly into the mathematics or not ramped beyond basic concepts at all. As such, it can be challenging to move from an interest in exploring quantum to research-level competence.

In this paper, we seek to provide a road map for IE/OR researchers who seek to navigate to emerging landscape of QOR. We aim to provide a schema of major areas of the field and to lay out a pathway to develop from a novice-level understanding to research. Throughout, we will also highlight the importance of the study of applied questions in tandem with theoretical work.  
Application is where quantum can benefit people: the heart behind engineering as a whole. Researchers in IE/OR, who regularly think about the on-going cycle of application and theory, are particularly well-suited to lead in this space.

The remainder of this paper is structured as follows. In Section \ref{s:QCBasics}, we present the basics of quantum computing. We discuss quantum software and quantum hardware's current capabilities and projected future developments in Section \ref{s:Hardware}. In Section \ref{s:QCOR}, we provide an overview of major areas of algorithms that are particularly relevant to IE/OR. In Section \ref{s:QCApp}, we present areas of applied QOR research. In Section \ref{s:techdev}, we discuss potential next steps for technical development. In Section \ref{s:Disc}, we discuss the role of IE/OR in quantum computing and conclude.

\section{Basics of Quantum Computing} \label{s:QCBasics}

In this section, we will introduce the units of information (qubits) used in quantum computers, key quantum features most relevant for QOR (superposition, entanglement, and interference), and the circuit model. 

\begin{figure}
    \centering
    \includegraphics[width=0.5\linewidth]{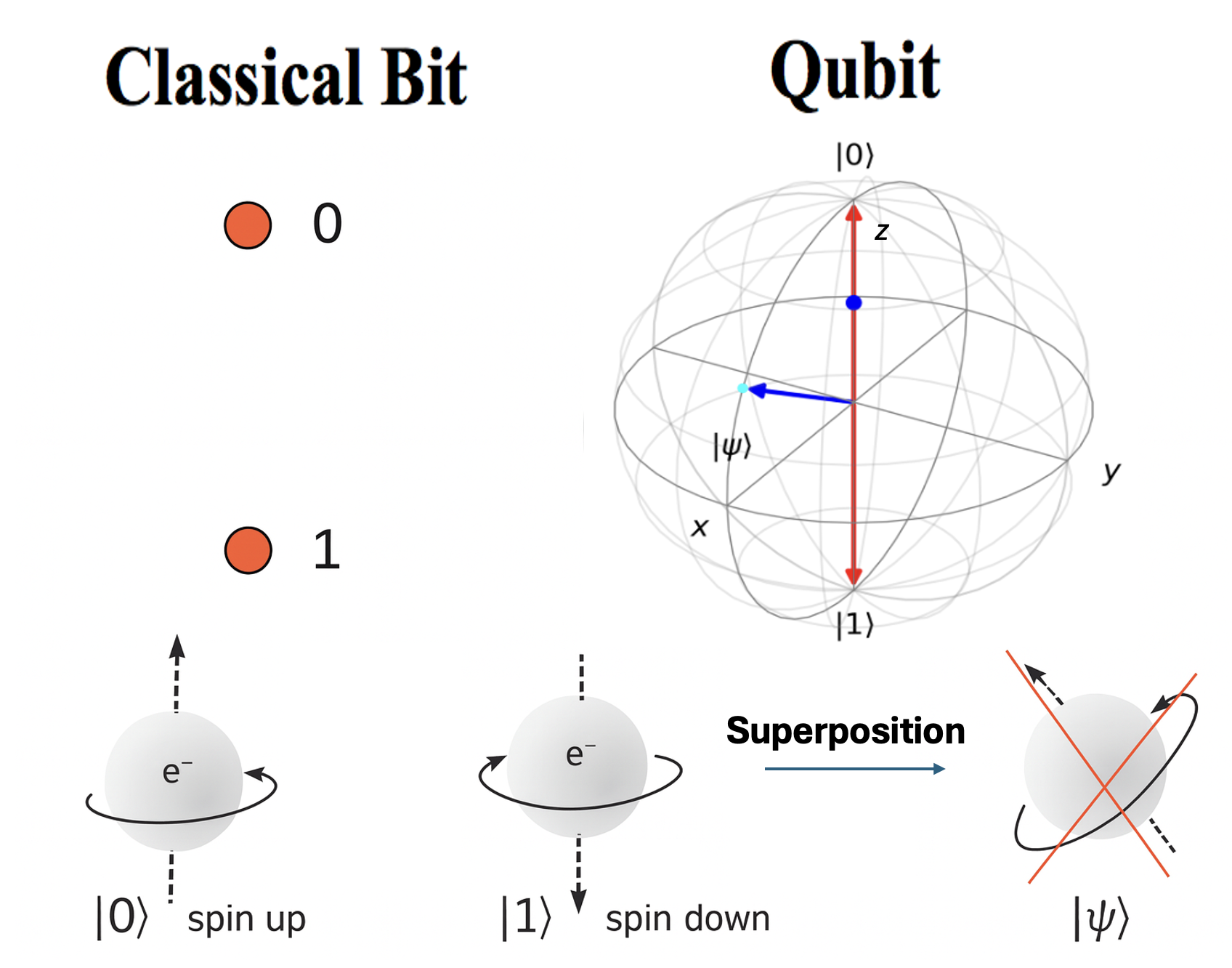}
    \caption{State of one qubit on Bloch sphere and visualization of spin electron for superposition.}
    \label{fig:qubit}
\end{figure}

Quantum computing uses quantum bits (qubits), which differ fundamentally from classical bits. A classical bit is either 0 or 1, but a qubit can exist in a \textit{superposition} of 0 and 1 states simultaneously (Figure \ref{fig:qubit}). Informally, a superposition means that when a qubit is measured it has a probability of being recorded in state 0 vs. state 1.

To be more formal, we will introduce the physics notation called ``Dirac" or bra-ket notation. The $| \psi \rangle$ (termed the bra, a column vector) and $\langle  \phi |$ (termed the ket, a row vector) indicate to the reader that $\psi$ is a qubit rather than a classical bit. The standard basis vectors are $|0\rangle$ and $|1\rangle$, termed spin up and spin down, respectively. They are represented in more familiar notation as $|0\rangle = [0, 1]^T$ and $|1\rangle = [1,0]^T$.

The state of a qubit is generally written as $|\psi\rangle = \alpha_0 |0\rangle + \alpha_1 |1\rangle$, where $\alpha_0$ and $\alpha_1$ are complex ``amplitudes" and $|\alpha_0|^2+|\alpha_1|^2=1$. When measured, a qubit collapses to one of the basis states $|0\rangle$ or $|1\rangle$ with probabilities $|\alpha_0|^2$ and $|\alpha_1|^2$, respectively. The state of a single qubit is described by a unit vector in a two-dimensional complex Hilbert space. 
For $2$ qubits, the quantum state can be expressed as
$|\psi\rangle = \alpha_{00}|00\rangle + \alpha_{01}|01\rangle + \alpha_{10}|10\rangle + \alpha_{11}|11\rangle, \text{ where } |\alpha_{00}|^2+|\alpha_{01}|^2+|\alpha_{10}|^2+|\alpha_{11}|^2=1$ 
which corresponds to a 4-dimensional complex unit vector. 

Note two benefits of these ideas. First, a system with $n$ qubits (called a quantum register) represents a unit vector in a $2^n$-dimensional complex space. This means the representational capacity grows exponentially with the number of qubits, i.e., 
quantum systems can encode  
information using exponentially fewer resources than classical systems. Second, the superposition principle also enables quantum computers to process a combination of many states at once, providing a form of parallelism beyond the classical bit.

Another key quantum phenomenon is \textit{entanglement}, which arises in systems of two or more qubits. An entangled multi-qubit state cannot be factored into independent single-qubit states. This means that measuring one qubit instantaneously influences the state of its entangled partner(s), no matter how far apart they are. Entanglement enables correlated outcomes and is a resource that quantum algorithms exploit to perform tasks that are hard for classical systems. A famous example is the 2-qubit Bell state $|\Phi^+\rangle=\frac{1}{\sqrt{2}}(|00\rangle+|11\rangle)$, in which both qubits are perfectly correlated. If one qubit in $|\Phi^+\rangle$ is measured and found to be $0$, the other immediately is known to be $0$ as well (and likewise for $1$), even though prior to measurement the joint state was a superposition of $|00\rangle$ and $|11\rangle$. Entanglement is non-intuitive but underpins quantum speedups in many algorithms.

\begin{figure}
    \centering
    \includegraphics[width=0.5\linewidth]{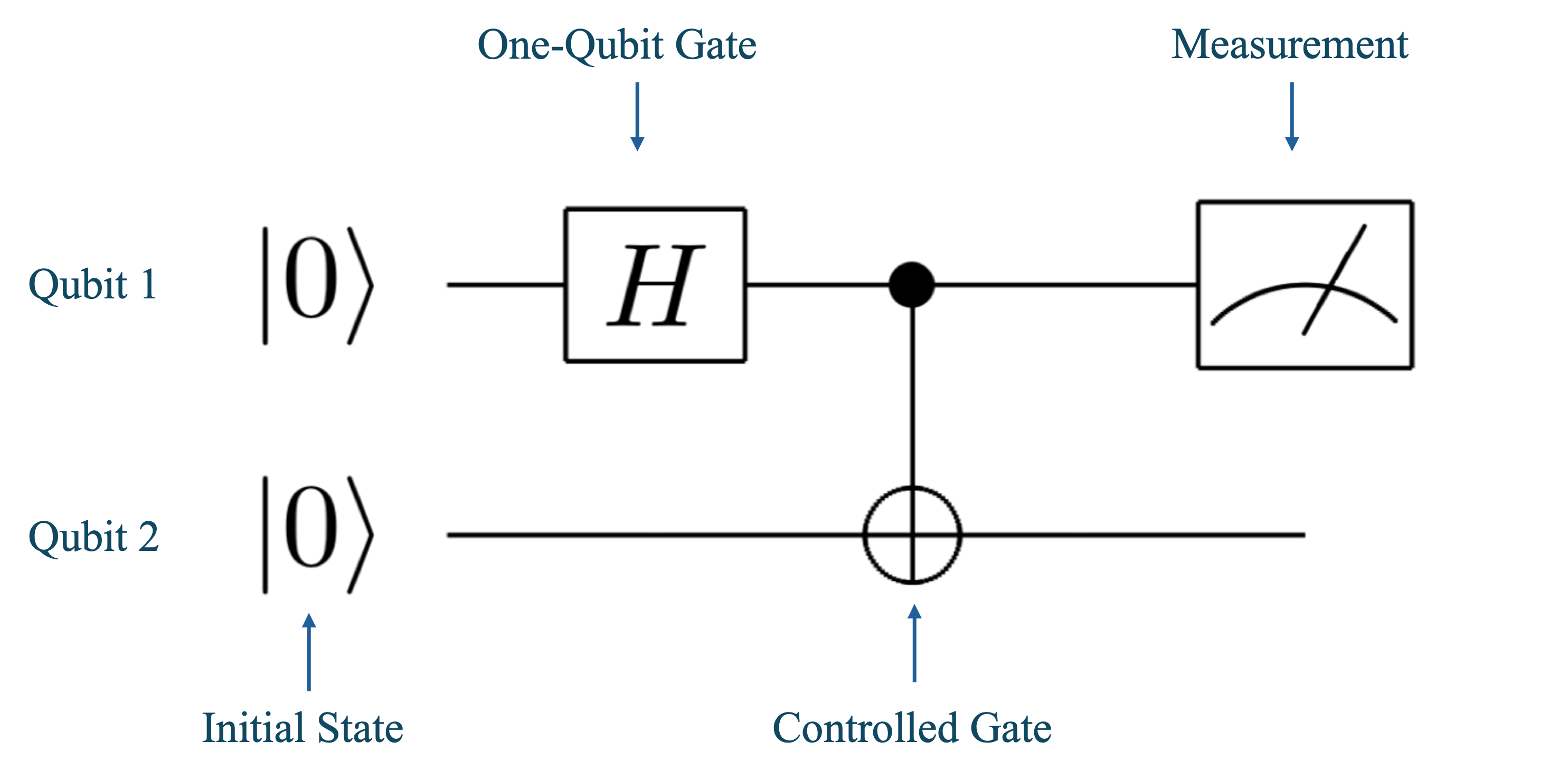}
    \caption{Quantum circuit.}
    \label{fig:circuit}
\end{figure}

Quantum algorithms are typically described using the \textit{quantum circuit model} (e.g., Figure \ref{fig:circuit}). Circuit diagrams are read left-to-right and represent how each qubit is operated on by quantum logic gates (i.e., matrix operations). These quantum gates are reversible linear transformations on the state vectors of the corresponding qubit(s); each gate may operate on one or more qubits. 

In Figure \ref{fig:circuit}, there are three qubits in the quantum register. Qubits 1 and 2 start in state $|0\rangle$. Qubit 1 is then operated on by the one qubit Hadamard (H) gate, i.e., $H|0\rangle = \frac{1}{\sqrt{2}}(|0\rangle+|1\rangle)$, as Hadamard gate builds a superposition. Note gates are left-multiplied on the current state vector(s). Then a two qubit gate (CNOT) entangles qubits 1 and 2. It is easy to verify that the resulting state is Bell state $|\Phi^+\rangle$. 
Whenever a measurement operator is applied, it collapses the qubit to either $|0\rangle$ or $|1\rangle$ based on the current state vector; this is irreversible and recorded on a classical bit (noted by double wires). These yield probabilistic classical outcomes.

Because quantum states can exhibit \textit{interference} (the amplitudes can add or cancel out like waves), quantum circuits can be designed so that the amplitudes of wrong answers destructively interfere and the amplitudes of correct answers constructively interfere. This ability to engineer interference is what lets quantum algorithms sometimes outperform classical ones.

The power of qubits, superposition, entanglement, and interference is perhaps best illustrated by a few landmark algorithms developed in the 1980s and 1990s:

  \textit{Deutsch’s algorithm} was the first quantum algorithm to demonstrate a speedup over classical computing. In the simplest case, one is given a hidden Boolean function $f:\{0,1\}\to\{0,1\}$ and asked whether $f(0)$ and $f(1)$ are the same or different. Classically, one would need to evaluate the function twice (for 0 and 1) in the worst case. Deutsch’s quantum algorithm uses a single quantum evaluation of $f$ (in superposition) to determine $f(0)\oplus f(1)$ with certainty. This algorithm, while solving a toy problem, introduced key ideas like querying a function on a superposition of inputs and using interference to extract global properties of $f$. It was later generalized to the Deutsch-Jozsa algorithm, which can distinguish constant functions from balanced ones using one query, whereas a classical deterministic algorithm would need multiple queries \citep{deutsch1992rapid}.
    
    \textit{Grover’s algorithm} 
    tackles the unstructured search problem: finding a marked item in an unsorted list of $N$ elements \citep{grover1996fast}. Classically, searching requires $O(N)$ queries in the worst case. Grover’s quantum search algorithm finds the target item with high probability in only $O(\sqrt{N})$ queries, providing a quadratic speedup. Grover’s algorithm repeatedly applies an amplitude amplification technique, essentially performing iterations of querying an oracle for the target and then inverting about the average amplitude, which increases the probability of the marked state linearly with each iteration. Grover’s technique is widely applicable as a subroutine and has been used in numerous other algorithms.

    \textit{Shor’s algorithm} 
    demonstrated an exponential speedup for a problem of practical significance: integer factorization \citep{shor1994algorithms}. Shor’s algorithm can factor an $n$-digit integer in roughly $O(n^3)$ time (more precisely, polynomial in $n$), whereas the best known classical algorithms run in sub-exponential but super-polynomial time. The key idea of Shor’s algorithm is a reduction of factoring to finding the period of a certain modular arithmetic function using quantum phase estimation. The algorithm employs quantum parallelism to evaluate the function on a superposition of inputs and uses the quantum Fourier transform to extract the period. Shor’s dramatic speedup highlighted the impact of quantum computing on cryptography: it breaks RSA encryption if a large fault-tolerant quantum computer is built. See Section \ref{s:Hardware} for a discussion of fault tolerance. This result spurred enormous interest in quantum computing research.

These algorithms rely on the ability of qubits to explore many possibilities at once and use interference to amplify correct answers. However, it is important to note that quantum algorithms do not simply ``try all solutions in parallel.” For instance, Grover’s algorithm does not brute-force check each item faster; rather, it amplifies the amplitude of the correct answer through structured interference across iterations. Similarly, Shor’s algorithm uses the structure of periodic functions and the Fourier transform to find factors faster than trial division. Designing quantum algorithms requires finding problems with exploitable mathematical structure or using quantum subroutines (like amplitude amplification or phase estimation) to gain advantages.

\section{Quantum Computing Software and Hardware} \label{s:Hardware}

There is a growing quantum software ecosystem that IE/OR researchers can use to program algorithms on simulators and actual hardware. 
Three major quantum programming frameworks implemented as Python libraries and SDKs: Qiskit, Cirq, and PennyLane (see Supplementary Materials for details). Access to quantum hardware and simulators is available through the hardware companies directly, cloud platforms that connect to many hardware vendors (e.g., Amazon Braket, Microsoft Azure Quantum, qBraid), and/or direct hardware access through your organization (e.g., if it has a quantum computer). 

Because near-term quantum devices are limited, simulation platforms and emulators play an important role. Classical simulation of quantum circuits is exponentially hard in the worst case (each additional qubit doubles the state space), but clever techniques and significant computational resources can simulate intermediate sizes. Researchers often test new quantum algorithms on simulators like the ones included in Qiskit (the Aer simulator) or Cirq, which can handle around 30 qubits on a laptop or more with HPC resources. There are also specialized simulators: for instance, tensor network simulators can sometimes simulate 50+ qubits by exploiting low circuit complexity, and emulators that inject noise models to mimic a particular hardware’s behavior. Simulation is invaluable for algorithm development and for understanding hardware needs before running on real machines. Additionally, companies have built analog quantum simulators for specific problems (such as cold atom analog simulators for physics). These are not gate-model computers but rather laboratory systems that directly mimic a Hamiltonian of interest (useful in studying material science or chemistry problems). For industrial engineers, simulators offer a way to validate quantum approaches on small instances of problems where results can be compared to classical benchmarks, all before deploying on an actual quantum device. Simulators also aid in developing error mitigation strategies by providing an environment where one can toggle noise on and off to see its impact on algorithm performance.

Regarding hardware, the current era of quantum computing hardware is often called the “NISQ” era, which stands for Noisy Intermediate-Scale Quantum. NISQ devices have on the order of tens to a few hundreds of physical qubits, and these qubits are prone to errors (noise) and have limited coherence times. Importantly, NISQ devices do not yet employ quantum error correction across all qubits, so computations are analog in nature and must complete before decoherence and noise overwhelm the result. Despite these limitations, NISQ machines are the first experimental platforms where quantum algorithms can be run and tested outside of simulation. For example, Google’s 53-qubit Sycamore processor and IBM’s 127-qubit Eagle processor are NISQ devices – they have achieved groundbreaking milestones like quantum supremacy demonstrations (performing a contrived computation much faster than a classical supercomputer).  
However, their operations are noisy, so algorithms must be short (shallow circuits) to have any chance of success before errors accumulate. Researchers are actively exploring how to get useful results from NISQ devices, via techniques like error mitigation and hybrid quantum-classical algorithms.

By contrast, the long-term goal is to build fault-tolerant quantum computers. A fault-tolerant device would employ quantum error correction to encode logical qubits into multiple physical qubits, in such a way that computations can be arbitrarily long with negligible error rates. Fault-tolerant quantum computers require many physical qubits—estimates often run into the millions, to create a handful of stable logical qubits capable of running deep algorithms like Shor’s factoring. While NISQ devices are measured by qubit count and gate fidelity, fault-tolerance will be measured by logical qubits and error correction overhead (e.g. the surface code requires on the order of 100 physical qubits per logical qubit, depending on physical error rates). The threshold theorem of quantum computing guarantees that if physical error rates can be pushed below a certain threshold, error correction can reduce logical error rates to arbitrarily low levels. Achieving this in practice is a major engineering challenge. Approaches like topological qubits (e.g. Majorana-based qubits) and low-error superconducting or ion-trap qubits are being pursued to reach the fault-tolerance regime. In summary: NISQ devices are our present reality – small, noisy quantum processors – whereas fault-tolerant devices represent the future of scalable quantum computing with error-corrected, reliable operations.

Current quantum hardware implementations use various physical technologies for qubits, each with pros and cons. The leading platform to date is superconducting qubits, used by IBM, Google, and others. These qubits are tiny circuits (Josephson junctions) on a chip that exhibit two energy levels that can serve as $|0\rangle$ and $|1\rangle$. Superconducting qubits are controlled by microwave pulses and typically housed in dilution refrigerators at millikelvin temperatures. They achieve gate operation speeds on the order of tens of nanoseconds, which is fast, but they suffer decoherence on the scale of tens of microseconds, limiting the depth of circuits that can be run reliably. Nonetheless, steady engineering progress has increased qubit counts (IBM plans a 1121-qubit device in 2025) and improved fidelities year over year. Superconducting devices have demonstrated small algorithms and even a form of quantum advantage in random circuit sampling.

Another prominent technology is trapped-ion qubits, used by IonQ, Quantinuum (Honeywell), and academic groups. Ion qubits use individual atoms (like Ytterbium or Calcium ions) trapped in electromagnetic traps and manipulated with laser beams. The qubits are internal electronic states of the ions. Trapped ions have the advantage of very long coherence times (seconds) and all-to-all connectivity (any ion can be entangled with any other via collective motional modes). High-fidelity gates have been achieved on small numbers of ions. The challenge with ions is slower gate speeds (microsecond to millisecond gates due to mechanical motion and laser switching) and the difficulty of scaling to large numbers of ions in one trap. Companies are exploring modular ion trap architectures to scale beyond 100 qubits by networking smaller traps. Other qubit technologies include photonic qubits (e.g. used by Xanadu’s optical quantum computers), neutral atoms (Rydberg atom arrays manipulated by lasers), and even semiconductor spin qubits (quantum dots, Si/SiGe spins). Each technology is at a different stage of maturity, but superconducting and trapped ions currently lead in demonstrating algorithms.

Beyond the circuit-model gate-based quantum computers, there is a distinct hardware approach known as quantum annealers. Quantum annealing devices, like those made by D-Wave Systems, are designed to solve optimization problems using a special-purpose quantum protocol. Instead of gates and circuits, an annealer operates by initializing qubits in a known easy state (often the ground state of a simple Hamiltonian) and then continuously evolving (annealing) the physical parameters so that the system’s Hamiltonian encodes the cost function of an optimization problem. If the process is slow enough and quantum fluctuations are present, the system will ideally stay in the ground state of the evolving Hamiltonian and end up in a state that encodes the optimal or near-optimal solution of the original problem. D-Wave’s machines, which now have over 5000 qubits, use superconducting flux qubits connected in a specific graph topology to realize an Ising model (a network of spins with tunable couplings). Users can program (embed) an optimization problem into the machine by setting the coupling strengths and local fields, and the quantum annealer finds low-energy states that correspond to good solutions. It is important to note that D-Wave’s qubits are highly connected but also very noisy; the annealing process is not fully error-corrected. 
Yet, quantum annealing is a valuable pragmatic approach; it is physically easier to engineer analog quantum fluctuations than a universal gate model with error correction, so D-Wave has been able to scale to thousands of qubits (albeit noisy, analog qubits). 
Initial experiments provide insight into how quantum effects (e.g., tunneling between many nearly-optimal solutions) might help in optimization. 

Given the limitations of current hardware, there is a strong push for co-design of algorithms and hardware. Co-design means developing quantum algorithms tailored to the realities of NISQ hardware, and conversely designing hardware features that make specific algorithms run more efficiently. For example, variational algorithms (discussed in the next section) are often designed to be shallow (few gate layers) to cope with short coherence times. Similarly, quantum error mitigation techniques use extra circuit runs and classical post-processing to cancel some effects of noise without full error correction, allowing NISQ devices to output more accurate results than raw hardware would allow. On the hardware side, new features like mid-circuit measurement and feedforward (available on some IBM devices) enable dynamic circuits where one can measure a qubit mid-computation and adapt later operations based on the result, which can improve the implementation of algorithms like quantum error correction or certain variational algorithms. Another co-design aspect is improving qubit connectivity or adding tunable couplers to better embed problem graphs for optimization algorithms. As an example, if a quantum chip has a connectivity graph matching a certain optimization problem’s graph, it can solve that problem more directly without needing many SWAP gates or complex embeddings. Researchers in operations research and computer architecture are collaborating to formulate requirements for next-generation hardware that would make near-term quantum algorithms more effective. This co-design philosophy is seen as crucial for bridging the gap between what algorithms assume (in theory) and what current devices can do.

In summary, quantum hardware has made great strides, but we are still in the early stages. NISQ devices provide a testing ground for quantum algorithms and a taste of quantum advantage in special cases, while significant engineering challenges remain to reach fault-tolerant scales. Nonetheless, even these early devices have enabled an ecosystem of experiments and prompted innovations (like variational algorithms) that are directly relevant to operations research and industrial engineering problems. Researchers in these fields can start engaging with quantum computing now—via cloud-accessible NISQ machines or high-performance simulators—so that they are ready to exploit fully error-corrected quantum hardware when it arrives in the future.
\section{Quantum Algorithms for IE/OR} \label{s:QCOR}
Quantum computing has the potential to impact many areas of IE/OR 
by offering new algorithms for mathematical problems at the heart of optimization, simulation, and machine learning. We organize this section into four subcategories that align with major OR topics: linear algebra, optimization, machine learning, and stochastic simulation. Throughout, it is important to emphasize an honest outlook: quantum algorithms often come with caveats (e.g., requiring fault-tolerant hardware or having complexity bounds under specific assumptions). The goal here is to outline what has been explored and what the theoretical speedups could be.

\begin{figure}[ht!]
    \centering
    \includegraphics[width=1.1\linewidth]{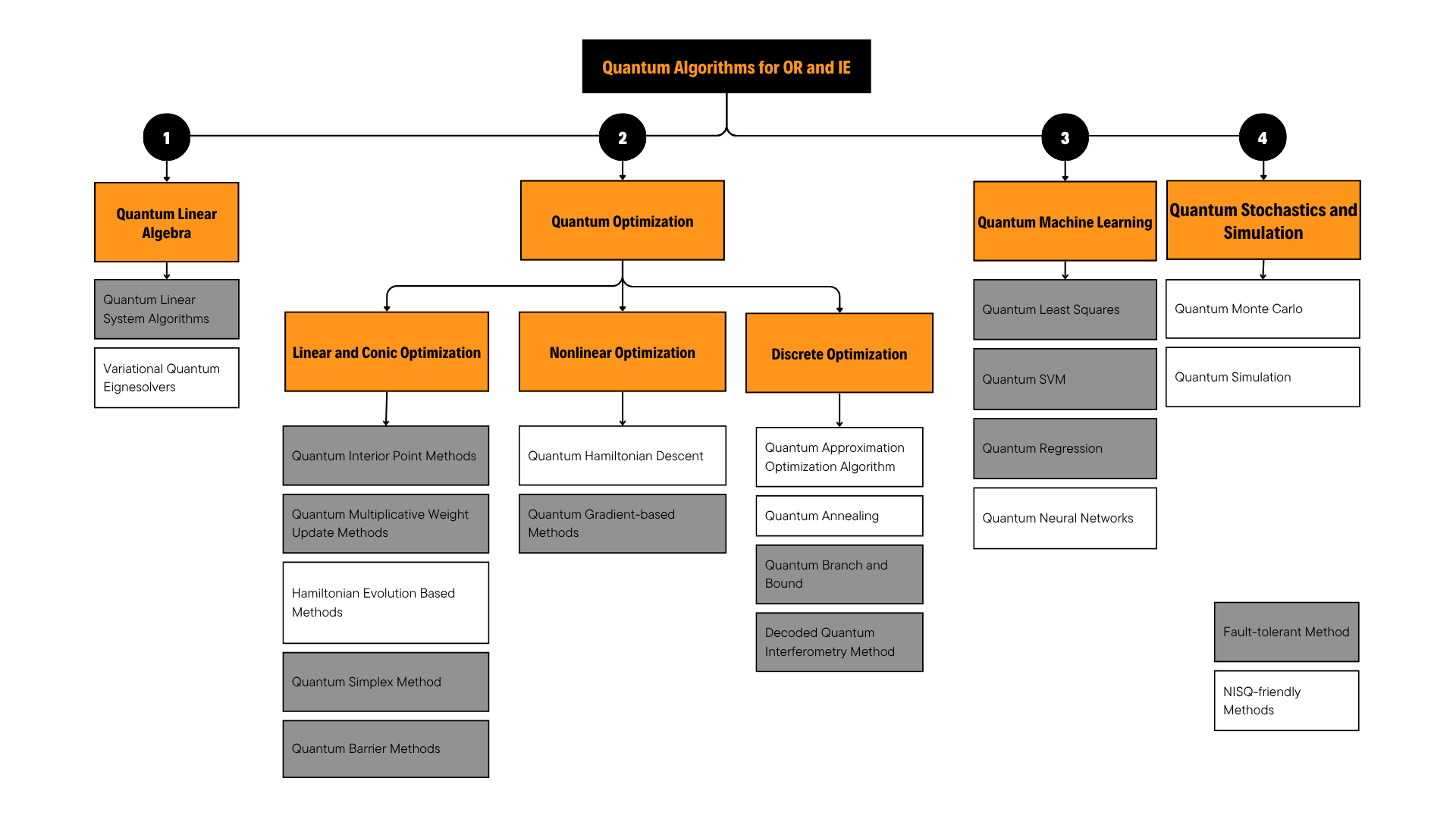}
    \caption{Quantum Algorithms for IE/OR.}
    \label{fig:algos}
\end{figure}
\subsection{Linear Algebra} \label{s:QLA}
Many OR problems involve linear algebra steps – solving systems of linear equations, eigenvalue estimation, matrix factorizations, etc. A cornerstone result in quantum algorithms is the HHL algorithm (named after its inventors Harrow, Hassidim, and Lloyd) for solving linear systems of equations exponentially faster than classical methods in certain conditions \citep{harrow2009quantum}. The 
algorithm, 
addresses the problem of solving $A\vec{x} = \vec{b}$ for $\vec{x}$, where $A$ is an $N\times N$ matrix and $\vec{b}$ is a known vector. Using quantum phase estimation, HHL can in principle find a quantum state proportional to the solution vector $|\vec{x}\rangle$ in time $O(\log N)$, an exponential improvement over the best classical solvers which take $O(N s)$ time for $s$-sparse $A$. However, this speedup comes with important constraints: HHL assumes one can efficiently prepare the right-hand side as a quantum state and efficiently implement Hamiltonian dynamics $e^{-iAt}$, and the runtime also scales with matrix condition number (i.e., ill-conditioned matrices negate the advantage). Moreover, the output of HHL is a quantum state $|\vec{x}\rangle$, so obtaining the full classical solution $\vec{x}$ would require further quantum-to-classical readout (which can be expensive, often reducing the practical speedup). Despite these caveats, HHL was a breakthrough as a proof-of-concept that certain linear algebra problems might be solved in time with at least polynomial speedup. This has inspired a line of research into quantum linear algebra solvers and simulators.

For industrial engineers, a relevant example is linear programming: interior-point methods for linear programs often require solving linear systems at each iteration. If those linear systems could be solved faster by a quantum subroutine, the overall optimization could be accelerated. Indeed, some proposals for quantum interior point methods (covered in Section \ref{s:QCOPT}) 
use quantum linear solvers as a subroutine. Another example is least-squares regression: one can formulate it as solving $(X^T X)\vec{\beta}=X^T \vec{y}$ for the coefficient vector $\vec{\beta}$. Quantum algorithms have been proposed to estimate regression coefficients faster than classical least squares, again under sparsity or low-rank assumptions. There are also quantum algorithms for eigenvalue problems, such as quantum principal component analysis (quantum PCA) which can project data into a principal component subspace without explicitly computing the covariance matrix eigen-decomposition. While these algorithms often assume access to data in quantum state form or in quantum random access memory (QRAM), they suggest the possibility of tackling high-dimensional algebraic problems in new ways. 

A particularly promising near-term approach in quantum linear algebra is using variational algorithms: instead of HHL’s deep circuit, one can set up a shallow parametrized circuit that encodes a candidate solution and then iteratively adjust parameters (with classical feedback) to minimize the residual $||A\vec{x}-\vec{b}||$. This falls under the umbrella of Variational Quantum Algorithms, and specific techniques like Variational Quantum Linear Solvers (VQLS) have been explored for small systems, showing that they can find solutions with fewer quantum resources by offloading work to classical optimization. While not offering exponential speedups, variational methods are more feasible on NISQ hardware and can handle some noise. This theme of hybrid quantum-classical algorithms will recur in the following sections.

\subsection{Quantum Optimization} \label{s:QCOPT}

Optimization is at the core of many industrial engineering problems, and it’s also one of the most active areas for quantum algorithms research. There are several quantum approaches to optimization and, as Figure~\ref{fig:QOPT} illustrates, for combinatorial and nonconvex problems, quantum algorithms aim better approximation ratio than classical approximation algorithms and for convex and conic optimization problems, quantum algorithms provide provable polynomial speedups to solve large-scale problems under some assumptions. We can categorize quantum optimization algorithms by the type of quantum hardware they target and whether they aim for exact or approximate solutions:
\begin{figure}
    \centering
    \includegraphics[width=0.8\linewidth]{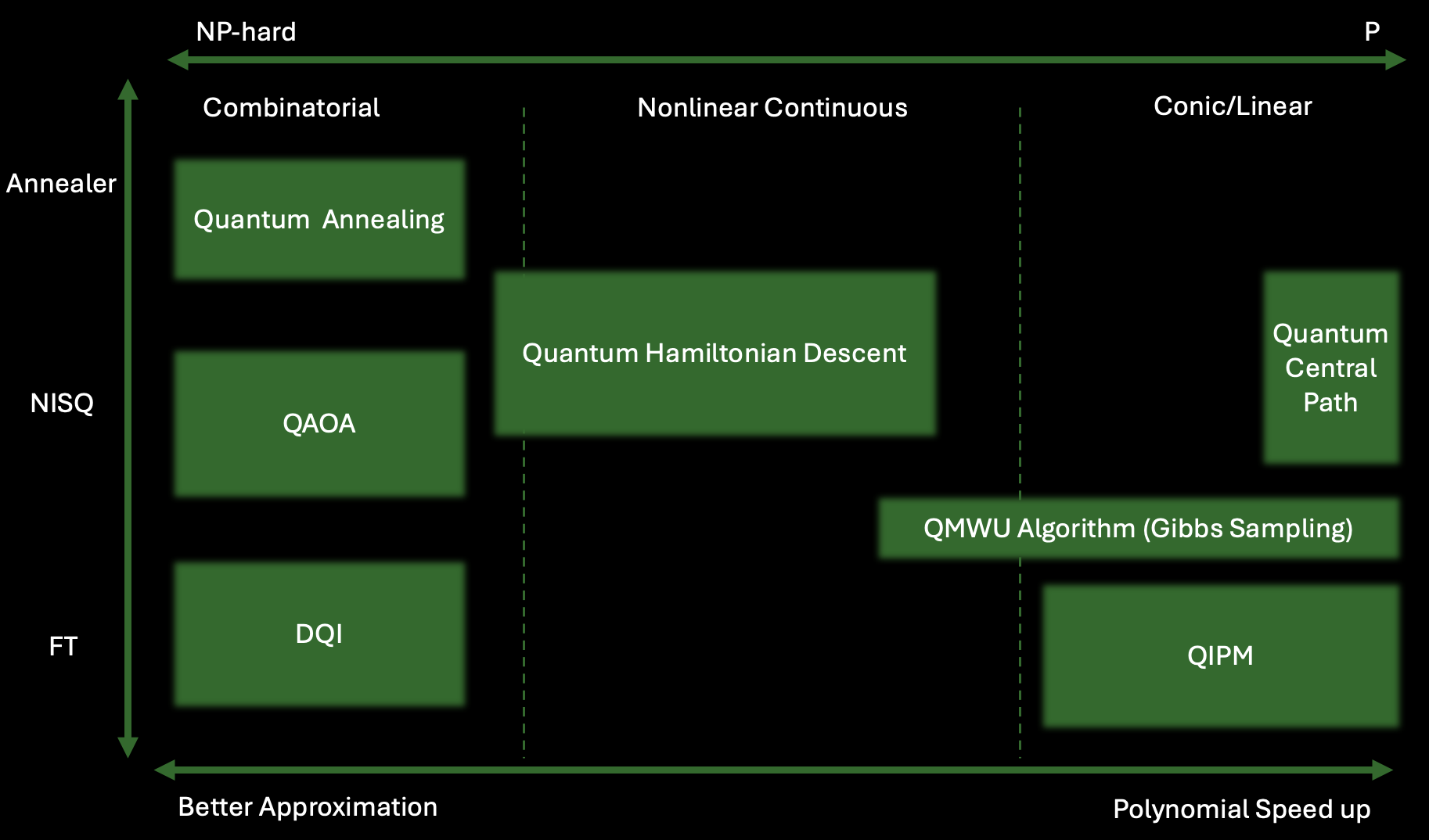}
    \caption{Optimization Problems and Quantum Algorithms to solve them.}
    \label{fig:QOPT}
\end{figure}
    
\textit{Quantum Annealing and Adiabatic Optimization}: As introduced in Section 3, quantum annealing is an analog method of optimization. The adiabatic quantum computing model is actually theoretically equivalent to the gate model in computational power—given enough time and coherence, an adiabatic evolution can perform the same computations as circuits \citep{albash2018adiabatic}. In practice, quantum annealers (like D-Wave) have been applied to various combinatorial optimization problems: scheduling \citep{perez2024solving}, traveling salesman\citep{martovnak2004quantum}, graph partitioning \citep{ushijima2017graph}, etc. The problem must be formulated as finding the ground state of an Ising spin glass or quadratic unconstrained binary optimization (QUBO) problem. This is very relevant to OR; many NP-hard problems can be mapped to QUBO form. While D-Wave’s machines have not definitively outperformed classical optimizers on real-world problems, they have sometimes found good solutions to specific instances and serve as a testbed for heuristic quantum optimization at scale (thousands of variables) that gate-model quantum computers cannot yet handle \citep{tasseff2024emerging}. Researchers have also developed quantum-inspired annealing algorithms (running on classical hardware) that mimic certain strategies of quantum annealing for large-scale problems, leading to cross-pollination between quantum physics and classical OR heuristics \citep{boev2021genome}.


\textit{Quantum Approximate Optimization Algorithm (QAOA)}: QAOA is a flagship gate-model algorithm for combinatorial optimization on near-term devices. Introduced by  \cite{farhi2014quantum}, 
QAOA is a variational algorithm that alternates between applying problem-specific operators and mixing operators on a set of qubits representing candidate solutions. For example, to solve a MAX-CUT problem, one can encode cuts in bitstring states of $n$ qubits, then use a cost Hamiltonian $H_C$ (derived from the graph’s adjacency matrix) and a mixing Hamiltonian $H_M$ (that flips qubit states) in an alternating sequence. With $p$ rounds (or depth $p$), QAOA produces a quantum state depending on $2p$ parameters; a classical optimizer is then used to tune these parameters to maximize the expectation of the cost $H_C$. At $p=1$, QAOA is basically a quantum variant of a single-step rounding algorithm; as $p$ increases, it can represent increasingly complex states (at $p \to \infty$, it can represent the optimal solution). 

\textit{Quantum Interior-Point and Other Exact Algorithms}: For fault-tolerant quantum computers (future devices with error correction), researchers have discovered algorithms that offer polynomial speedups for certain optimization problems in theory. One example is a quantum interior-point method for linear programming and semidefinite programming by \cite{kerenidis2020quantum}. 
Although QLSAs has the potential to speed up IPMs but it introduces new challenges like noise and inexactness of quantum subroutine can impact convergence of IPMs and for degenerate linear programs condition number of Newtons system goes to infinity making QLSAs inefficent. To address this issues, several techniques have been applied like refomulating Newton systems \citep{mohammadisiahroudi2025inexact}, using iterative refinement to boost precision \citep{mohammadisiahroudi2024efficient} and precoditioning techniques \citep{mohammadisiahroudi2025improvements} to achieve the optimal worst-case complexity $O(n^2L)$ up to some polylogarithmic factors under some assumptions like access to QRAM \citep{purdueai}. Similarly, quantum algorithms for semidefinite programs (SDPs) have been proposed that run in polynomially better time in the matrix dimensions, with additional dependence on solution precision and condition numbers \citep{augustino2023quantum}. These algorithms often rely on quantum linear algebra subroutines and block-encoding techniques (a method to load matrices into quantum operators). While promising, it is important to stress that these algorithms require fault-tolerance due to the large number of operations and the need for QRAM to input data. Nonetheless, they indicate that in principle, quantum computers can tackle certain continuous optimization problems faster, for example \cite{brandao2017quantum} proposed a non-IPM quantum algorithm based on the Multiplicative Weight Update Method (MWUM) of \citep{arora2012multiplicative} to solve Semidefinite Optimization (SDO) problems. 
After this paper, many improved versions of Quantum MWUMs (QMWUMs) are proposed for SDO \citep{van2018improvements, van2017quantum}, and LO \cite{van2019quantum}. 
The advantage of QMWUMs is that their complexity has a linear dependence on the dimension for SDO and a sublinear dependence for LO.
The major issue with QMWUMs is that they are highly dependent on inverse precision and an upper bound for the norm of the optimal solution, which in most cases, is exponentially large for~LO. 
However, this issue is not unique to QMWUMs, and the complexity of other quantum optimization methods, including the proposed approach of this paper, is dependent on an upper bound on the norm of the optimal solution directly or hidden in other parameters like the condition number bounds. 

For discrete optimization, apart from Grover’s search (which gives a generic quadratic speedup for brute force search), there are quantum algorithms that provide more nuanced advantages. For example, there are quantum approximate algorithms for specific NP-hard problems that sometimes beat the best known classical approximation ratios. For example, the algorithm Decoded Quantum Interferometry (DQI) \citep{jordan2024optimization} uses quantum Fourier transforms and decoding‐style reductions to give, for certain instances of problems such as max-XORSAT, better approximation ratios than any known classical polynomial‐time algorithm. These results are technical, but one takeaway is that quantum computers might not solve NP-hard problems in polynomial time (that remains highly unlikely), but they could give better approximate solutions or speed up certain heuristic search steps relative to classical methods. 

While much of the existing literature focuses on linear and convex optimization, nonlinear optimization represents the next frontier for quantum algorithms. Gradient-based methods are a natural starting point, as many nonlinear optimization techniques—such as gradient descent, Newton’s method, and quasi-Newton methods—rely on efficient computation of gradients or Hessians. Quantum algorithms can accelerate these computations through quantum differentiation \citep{jordan2005fast}. Beyond gradient estimation, more quantum-native optimization frameworks have emerged, notably Quantum Hamiltonian Descent (QHD), which generalizes classical gradient flow to the quantum domain \citep{leng2023quantum}. In QHD, the optimization trajectory is governed by the Schrödinger equation with a Hamiltonian constructed from the objective function, allowing quantum evolution to mimic gradient descent in the continuous limit. This formulation integrates naturally with quantum mechanics—energy minimization corresponds to objective minimization—and may leverage quantum coherence and tunneling to escape local minima. Together, quantum gradient estimation and Hamiltonian descent outline a pathway toward scalable nonlinear quantum optimization, connecting the structure of quantum mechanics with the geometry of optimization landscapes.

In the context of industrial engineering, a lot of attention has been paid to applying these quantum optimization methods to familiar OR problems: scheduling, vehicle routing, supply chain optimization, portfolio optimization, etc.. So far, quantum approaches have not surpassed classical operations research algorithms for practical problem sizes. For instance, classical mixed-integer programming solvers or specialized heuristics often find better solutions faster for small instances than current quantum devices or simulators can. However, the field is very much in an exploratory phase. Benchmarking efforts, such as the Quantum Optimization Benchmarking Library \citep{Koch2025}, are underway to fairly compare quantum and classical methods on representative problem sets. These will help identify niches where quantum heuristics might shine (e.g., problems with certain structures like high degeneracy or many local minima where quantum tunneling might help escape, as hypothesized in physics-inspired models). Even in the NISQ era, formulating OR problems for quantum solvers often yields new insights, sometimes leading to improved classical algorithms too. The pursuit of quantum optimization is thus a twofold win: we prepare for future quantum advantage and deepen our understanding of optimization problems and algorithms in general.

\subsection{Quantum Machine Learning} \label{s:QML}

Machine learning (ML) in industrial engineering spans data analysis, predictive modeling, clustering, and more. Quantum machine learning (QML) is the endeavor to use quantum computers to accelerate or improve machine learning tasks. There is considerable hype in this area but also a growing body of careful research clarifying what quantum computers might do for ML. Broadly, QML algorithms can be grouped into two types: (1) quantum speedups for linear algebra subroutines underlying ML (which overlaps with Section \ref{s:QLA}), and (2) using quantum states and operations as model components (like quantum neural networks or quantum kernels).

One of the earliest QML proposals was for quantum support vector machines (QSVMs). In classical SVM, one often needs to solve a quadratic optimization (for the soft-margin SVM) or to compute inner products in a high-dimensional feature space (kernel SVM). A quantum version by \cite{rebentrost2014quantum} leveraged a quantum linear system algorithm to potentially solve the SVM optimization faster in the case of very high-dimensional data encoded in quantum amplitudes. The idea is that if you can prepare a quantum state representing the data matrix, you might use HHL to find the SVM decision hyperplane with complexity polylogarithmic in the number of features. Similarly, quantum linear regression algorithms were proposed \citep{mohammadisiahroudi2022quantum}. These algorithms show exponential speedups in theory, but require the ability to load classical data into quantum states efficiently (the QRAM model). If data is dense and unstructured, this loading can nullify the quantum advantage. That said, for certain structured data or if one is given data in quantum form already, QSVM could be beneficial.

Another line of QML research involves quantum kernels and feature maps. A quantum computer can generate states that correspond to non-classical feature mappings of input data. By measuring overlaps between quantum states, one can compute kernel values that might be hard to compute classically. A notable experiment by \cite{Havlicek2019}
demonstrated a quantum classifier where the kernel is implicitly computed by a quantum circuit encoding the data, and this was shown to be intractable to simulate for a classical computer under plausible assumptions, hinting at a potential quantum advantage in classification tasks. Whether these quantum kernels actually give better classification accuracy on real-world data, or just equal accuracy attained in a faster way, is an open question. But it is an example of how quantum states can represent complex distributions or correlations that might be useful in ML. 

The most NISQ-friendly approach to QML is using parametrized quantum circuits as quantum neural networks (QNNs) \citep{benedetti2019parameterized}. These are sometimes called variational quantum classifiers or quantum neural nets. Essentially, one sets up a circuit with adjustable rotation angles (parameters) and possibly some entangling gates, and then trains those parameters via a classical optimization loop to make the circuit outputs match desired labels or predictions. This is analogous to a classical neural network where the quantum circuit plays the role of the network architecture and the measurements give you the model’s predictions. QNNs have been tested on small-scale problems like classifying simple datasets or as components in hybrid networks (for example, a quantum layer in a larger classical network). They have advantages in that the number of parameters can be much smaller than a classical network with similar expressive power, because a single quantum gate can produce very complex high-dimensional transformations of input data (thanks to superposition and entanglement) \citep{schuld2021effect}. However, training QNNs comes with challenges: as mentioned, the barren plateau problem is when the gradient of the cost function vanishes exponentially with the number of qubits, making training impractical. This tends to happen for random or too-deep circuits. Researchers are investigating how to design QNNs (perhaps with structured initializations or layered architectures) to avoid barren plateaus. They are also exploring advanced optimization techniques (like quantum-aware optimizers, second-order methods, or meta-learning approaches) to improve training convergence. Robust training of QNNs is an area where classical optimization (a strength of OR) can contribute techniques – for instance, derivative-free optimization or global optimization methods might be more appropriate than naive gradient descent for these tough landscapes.

In terms of applications, QML has been suggested for a variety of tasks: quantum clustering algorithms, quantum recommendation systems, quantum-enhanced genetic algorithms, etc. For example, there is a quantum version of $k$-means clustering that uses amplitude amplification to potentially speed up distance computations, and quantum PCA for dimensionality reduction \citep{mari2020transfer}. Another interesting direction is quantum generative models: quantum circuits (especially ones related to the structure of Boltzmann machines or normalizing flows) can represent probability distributions that might be expensive to sample classically. A quantum Boltzmann machine or quantum GAN could, in principle, generate data according to distributions that capture complex correlations (there have been experiments with simple quantum circuits generating toy images or distributions). These ideas remain at early stages; often a classical ML algorithm is superior on the small benchmarks we can run. Nonetheless, if large quantum computers become available, QML could handle massive feature spaces or complex models in ways classical computers cannot, by exploiting the exponentially large state space of $n$ qubits (which corresponds to $2^n$ complex amplitudes). A caveat is that having a state which implicitly contains $2^n$ numbers is only useful if we can also efficiently extract information from it (since reading out all those amplitudes is impossible). Thus, QML often focuses on tasks where the desired output is succinct (like a classification decision or a sample from a distribution), so that the quantum advantage can be realized in the end-to-end process.

Finally, it is worth noting that QML might not only be about speed. Some researchers ask if quantum models could generalize differently than classical ones, or be more robust to certain types of noise, etc. One often-cited motivation is representational efficiency: a quantum system of $n$ qubits lives in a $2^n$-dimensional state space, so a quantum model might represent certain functions or patterns with exponentially fewer resources than a classical model. However, whether this leads to better real-world ML performance is unproven – it could be that classical neural nets already exploit the necessary structure well. Current reviews urge caution and clarity in QML claims: for instance, \cite{schuld2022quantum} argue that chasing “quantum advantage” in ML might be less fruitful than understanding how quantum and classical can work together, and Cerezo et al. (2022) outline many challenges like trainability issues and noise that need to be overcome for QML to be useful.

For industrial engineers, QML could eventually enable faster processing of the huge volumes of data now common in operations (e.g. sensor data in manufacturing or transportation networks). It might also enable new kinds of simulation-based optimization or scenario analysis by efficiently sampling from complex probability distributions (blurring into the next section on simulation). In the nearer term, formulating ML problems in a way amenable to quantum computing often provides new angles, for example, thinking of data encoding as part of model design, or integrating domain knowledge as quantum oracles, which can sometimes feed back into improved classical methods or at least a deeper understanding of the problem structure.
\subsection{Stochastic Modeling and Simulation}\label{s:QSim}

Stochastic modeling, Monte Carlo simulation, and uncertainty quantification are bread-and-butter techniques in operations research, used in areas like supply chain risk analysis, financial engineering, reliability engineering, and simulation-based optimization. Quantum computing offers some intriguing algorithms that could accelerate Monte Carlo simulation and related stochastic computations by leveraging amplitude amplification techniques.

The core idea is embodied in the quantum amplitude estimation (QAE) algorithm. Classical Monte Carlo typically involves sampling a random variable or system many times to estimate an expectation or probability to some desired precision. To estimate an output with standard error $\epsilon$ usually requires on the order of $O(1/\epsilon^2)$ samples classically (because the error decreases as $1/\sqrt{N}$ after $N$ samples, by the central limit theorem). Quantum amplitude estimation, introduced by Brassard et al. and refined by others, can achieve the same precision $\epsilon$ with quadratically fewer samples, $O(1/\epsilon)$. It does this by essentially using quantum phase estimation on the amplitude of a “good” outcome to more precisely estimate that amplitude than would be possible by simple sampling. In practical terms, if you have a quantum subroutine (oracle) that recognizes when a simulation outcome is a success (or tallies a payoff), QAE allows you to amplify the amplitude corresponding to that success and measure it to obtain a high-precision estimate of its probability. This was a major theoretical discovery because it shows quantum computers can in principle quadratically speed up a very general class of computations – estimating integrals and expected values, which covers Monte Carlo simulations used in finance, project management (e.g. schedule risk analysis), and more.

To ground this with an example: consider valuing a financial derivative by Monte Carlo simulation. One needs to simulate many random price paths and take an average payout. With quantum amplitude estimation, one would prepare a quantum state that encodes the random path generation and payout calculation, then use QAE to estimate the average payout with fewer runs than classical Monte Carlo. This approach has been studied for option pricing and risk analysis in finance. In fact, a recent study on quantum computing for finance highlighted that amplitude estimation could speed up Value-at-Risk or option pricing calculations asymptotically. The challenge is that current hardware cannot implement QAE directly for practical problems – it requires long coherent circuits (quantum phase estimation is not NISQ-friendly) and the advantage only shows up when extremely high precision is required (because the constant factors in quantum circuits are large). However, researchers have devised variations like iterative QAE or Bayesian QAE that try to be more NISQ-compatible (by avoiding the need for a large quantum Fourier transform) at the cost of some additional circuit runs or post-processing. Even these variants, when tested on small examples, confirm the quadratic convergence behavior. So, the theory is solid: as soon as we have moderately large, low-noise quantum computers, amplitude estimation could become one of the first useful tools for industry, accelerating simulation tasks.

Another area of interest is quantum generation of random samples or distributions. For example, generating true random numbers via quantum processes is already used in some high-security applications (quantum random number generators). But more interestingly, quantum computers can prepare certain probability distributions more efficiently than classical methods. A prime example is preparing a Gibbs distribution (Boltzmann distribution) of a complex system \citep{amin2018quantum}. Quantum algorithms can prepare a quantum state corresponding to a Gibbs distribution at temperature $T$ potentially faster than classical sampling would mix to that distribution. This has implications for things like sampling-based optimization (quantum simulated annealing in the gate model) or for probabilistic graphical models in machine learning. If one could sample from a complicated high-dimensional distribution (say, the stationary distribution of a Markov chain) more quickly, it could speed up algorithms in operations research that rely on such sampling. One quantum algorithm for this is through quantum Markov chain Monte Carlo or quantum walks, which under some circumstances can mix quadratically faster than classical Markov chains. These are advanced topics, but the key point is that quantum simulation isn’t only about simulating quantum physics – it can also aid in simulating classical stochastic systems by using quantum processes to speed up random sampling.

For more direct simulation of physical systems (as originally envisioned by Feynman), quantum computers excel at problems like chemical or materials simulation, which may not seem directly tied to OR but can be in industrial domains (e.g., simulation-based optimization of chemical process yields, battery materials design, etc.). Industrial engineers working in fields like semiconductor manufacturing or material design might find that quantum simulation algorithms allow them to analyze models that are intractable classically, thus providing inputs (e.g., material properties, reaction rates) into larger OR models that were previously unavailable.

One more stochastic application is in queuing networks and random processes. There is research into quantum algorithms for rapidly mixing random walks or solving linear systems that arise in Markov chain steady-state analysis. For example, computing the steady-state of a large Markov chain (common in queuing analysis) can be framed as solving linear equations $(I - P^T)\pi = 0$. Quantum linear solvers (like HHL) could in principle find the stationary distribution $\pi$ faster for certain types of transition matrix $P$, especially if $P$ is sparse. Also, quantum walk algorithms can sometimes detect properties of graphs (like connectivity, centrality measures) faster than classical algorithms, which might be of interest in network optimization and analysis.

It is worth noting that while amplitude amplification gives a quadratic speedup in sample count, the overall speedup for a complete application can be less if a lot of classical post-processing or problem-specific overhead is involved. Also, setting up the quantum oracle (the subroutine that flags a ``success" in amplitude estimation) can be non-trivial and may require circuits as complex as the simulation itself. This is an active area of development: finding clever ways to construct quantum oracles for complex events or payoffs that don’t outweigh the quadratic advantage.

From a high-level perspective for industrial engineering: if one needs to estimate a KPI (key performance indicator) that is defined as an expectation over many random scenarios (for instance, average cost of a stochastic schedule, risk of a rare event in a supply chain, etc.), a quantum computer could in the future provide that estimate with far fewer runs than today’s Monte Carlo. In optimization under uncertainty, this could translate to faster evaluations of objective functions or constraints that are expectations, thereby accelerating algorithms like simulation-based optimization or stochastic programming (by reducing sample counts in each iteration). For rare-event simulation, amplitude amplification is particularly enticing, because if an event has probability $p \ll 1$, classical simulation needs $\sim 1/p$ samples to get even a single occurrence on average, whereas quantum methods could in principle amplify that probability with quadratically fewer tries.

Quantum algorithms for stochastic simulation are one of the more practically promising avenues once hardware matures: they offer clear mathematical speedups that apply to generic and widely-used computational tasks in OR/MS. Demonstrating these advantages experimentally will require fault-tolerant quantum computers with dozens or hundreds of logical qubits, which are still some years away. Until then, researchers continue to test small instances on today’s devices and refine algorithms (for example, combining amplitude estimation with error mitigation, or using hybrid quantum-classical approaches to handle noise). The prospect of solving in minutes what would classically take days of Monte Carlo simulation is a powerful motivator for continuing research in quantum computing for stochastic modeling.

\section{Quantum Computing for Applications}\label{s:QCApp}

The area of applied QOR focuses on the connection between quantum OR methods and real-world problems. Much like in classical OR, where there is a loop between new methods driving applied impact, which motivates new methods which open up new applications and so forth, we expect the same looping structure to occur with quantum OR. As new QOR methods are developed, it may be possible to solve new types of applied problems, and new applied problems may motivate new quantum methods.

\begin{figure}[ht!]
    \centering
    \includegraphics[width=0.7\textwidth]{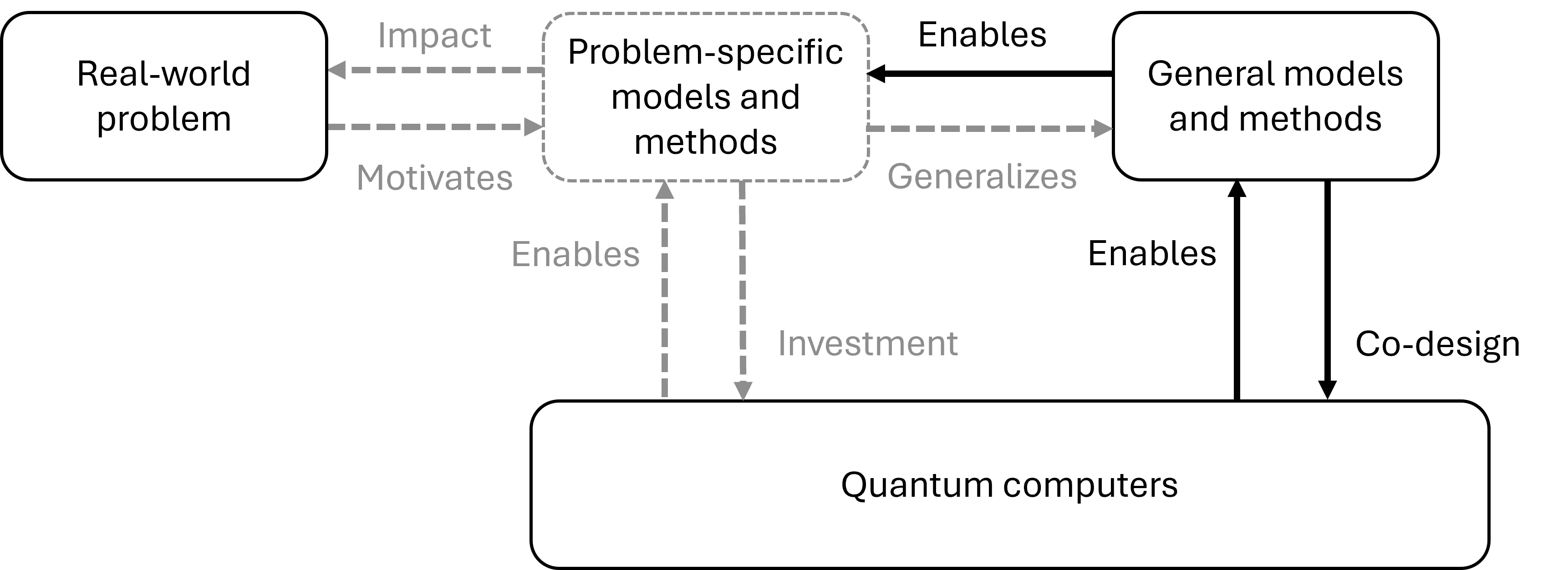}
    \caption{The Role of Applied OR in a Quantum Context}
    \label{Fig:appliedOR}
    \vspace{-10pt}
\end{figure}

The cycle of applied and generalized quantum OR is illustrated in more depth in Figure \ref{Fig:appliedOR}. What is unique about quantum OR (relative to classical) is that the underlying hardware development is tied into this loop. As quantum computers develop, we are able to solve new types of problems and develop new methods (Figure \ref{fig:algos}). Beyond this, though, the development of quantum computers is affected by progress in quantum OR: through co-design (Section \ref{s:Hardware}) as well as in investment. It is the potential use cases of quantum computers that drive large investments in hardware development.

In this section, we will discuss research and practice of applied quantum OR (designated in grey on Figure \ref{Fig:appliedOR}). We first turn to the types of research contributions under active study. One key area within applied OR is modeling, i.e., the translation of a decision problem in a corresponding mathematical expression that can be solved via algorithms developed for quantum hardware. General quantum OR models are those that can be applied to a range of problem types. As noted in Section \ref{s:QCOPT}, one popular style now is Quadratic Unconstrained Binary Optimization (QUBO) \citep{Kochenberger2006}. QUBO models are well-suited for quantum approaches because we can directly map the objective function to an Ising Hamiltonian that can be expressed with quantum circuits. A seminal presentation of NP-complete problems expressed as QUBOs is provided by \cite{Lucas2014}, with many specific models discussed in \cite{Glover2022}. We note that MaxCut problems are equivalent to QUBOs \citep{Barahona1989,Ivanescu1965}. Research on higher-order interactions is also on-going \citep{Babbush2013}, termed Higher Order Unconstrained Binary Optimization (HUBOs). One method is to use/develop ``gadgets” to express higher order interactions with quadratic terms, i.e., converting to a QUBO. The trade-off is that the number of decision variables grows polynomially with the number of terms of order three.

Within the area of problem-specific models and methods, the goal is to develop approaches to solve a given problem type better by some metric (e.g., faster, to a higher accuracy, with fewer resources). Much of the emerging quantum modeling research focuses on reformulating existing classical OR models into QUBO-style problems to allow them to be solved on quantum hardware. As an example, the quantum vehicle routing problem (VRP) literature is growing quickly \citep{Azad2023,Mohanty2023}.  One direction being explored is how to tailor abstracted VRP problems to specific settings. For example, to consider heterogenous vehicle capacities, \cite{Fitzek2024} develop an approach to integrate the traveling salesman and knapsack QUBO formulations originally presented separately in \cite{Lucas2014}. 
Another direction is to provide formulations that reduce the number of qubits needed (i.e., more efficient encodings). For example, \cite{Palackal2023} improves the CVRP formulation from \cite{Feld2019}. In the NISQ era, and moving forward, parsimonious formulations in terms of number of variables will be critical.

Models and methods are often intrinsically intertwined in applied OR, and in quantum applied OR, research is active in how to improve different steps within problem-specific methods. Given the small number of available qubits, problem size reductions are valuable. Researchers can lean on existing literature for preprocessing QUBOs \citep{Boros2006} as well as developed problem-specific approaches, e.g., eliminating variables that are known to be infeasible. It is not yet well-established what types of decomposition approaches work well for specific problems. Research is on-going for specific problem types (e.g., \cite{Herzog2024} evaluate the performance of a general quantum decomposition approach (wire cutting) to the Traveling Salesman Problem. 

As QUBOs and HUBOs are less commonly used in applied classical OR, their emphasis in quantum optimization may motivate new directions for expressing real-world problem domains. Further, be aware that when concepts from classical OR are ported to their quantum counterpart, the behavior may be different (e.g., problems with a Conditional Value at Risk (CVaR)-style objective may converge faster using variational methods than with expectation objective \citep{Barkoutsos2020,Kolotouros2022}). Recall as well that quantum models may be able to one day incorporate parameters that would be intractable classically, e.g., simulations of quantum systems (Section \ref{s:QSim}). Be creative; beyond performance improvements for problems we can currently solve classically, the impact of quantum computing for industrial engineering may well come from decision problem that are currently intractable, which will lean on applied QOR to find.

Research is also ongoing to study how to tailor general algorithms (Section \ref{s:QCOR}) to problem domains. These include the study of which mixer Hamiltonians perform well for particular problems and features that drive difficulty of problem instances (e.g., \cite{Brandhofer2023}). There may not be a clear winner on which algorithm will perform best for a given problem \citep{Mugel2022}, as stakeholders may vary in which metric (e.g., objective value, risk, solution time, cost to run) they choose to prioritize. As in classical models, it is important to understand and communicate these trade-offs to decision-makers, but note the quantum learning curve applies to stakeholder colleagues as well.

Key applied OR concepts of maintaining trust between the modelers, model, and stakeholders and ensuring model representativeness \citep{Harper2021} hold true in the quantum context as well. The varying degrees of stakeholder familiarity with quantum computing will be relevant in how to scope problems and how to communicate the results.
For problem scoping, it is crucial not to over-sell the near-term potential; some cutting-edge QOR problems can still be solved by hand, and we may still be years out from beating classical performance. Yet, figuring out what the practical near-term value is for the stakeholders is also worthwhile; remember that value goes beyond computational performance. For a given stakeholder, success may include supporting research, evaluating quantum as a strategic business direction, prestige, or something else entirely. The results should be then presented in way that will help them achieve that goal.

\section{Technical Development}\label{s:techdev}

The question then becomes how to progress from reading this paper to conducting meaningful work in quantum IE/OR. 
We consider three potential goals for IE/OR researchers and practitioners with respect to quantum: quantum-aware, a quantum IE/OR collaborator, and a quantum IE/OR professional. Quantum awareness is the ability to consider the potential near- and long-term benefits and limitations of quantum computing, specifically as it relates to IE/OR methods. It can be useful for individuals as they consider strategic directions as well as for those in leadership positions as they consider whether and the degree to which to invest in projects, centers, and hiring lines. A quantum IE/OR collaborator is able to collaborate with experts in quantum computation to provide technical expertise on IE/OR methods. For this outcome, the IE/OR professional would need to know enough about quantum computing to be able to participate in discussions but would not need to know how to development and implement the quantum methods themselves. This has the potential to accelerate the research in quantum computing related to IE/OR that is actively being conducted conducted by colleagues in other disciplines, e.g., physicists working on quantum discrete optimization. A quantum IE/OR professional is one who can advance quantum IE/OR applications and theory, including development, implementation, and practice.

In framing these three outcomes, we make a handful of observations. The first is that the outcomes are fairly sequential. It is possible to ``dip one's toes in" to explore the potential of quantum IE/OR without committing massive resources or time. The second is that the choice of where to focus relates to a perennial question in interdisciplinary research, i.e., do you contribute to an interdisciplinary team via your own disciplinary methods in collaboration with others or do you pursue expertise in multiple disciplines (here: both IE/OR methods and quantum computation). It is up to each professional to decide. 
Finally, the quantum ecosystem is developing rapidly, and there is the push for a further abstraction of the implementations, e.g., workflows to enable programmers to use quantum computers without programming circuits directly (e.g., \cite{Eichhorn2025}). While it is currently necessary to know about quantum computation to be able to implement models and algorithms, 
this may not always be the case. It may not be too far in the future that quantum computation becomes more accessible to those who aim for the quantum-aware and IE/OR collaborator outcomes as well. 

Next, to support development towards these outcomes, we present Table \ref{tbl:pathways}. The table provides a pathway of readings and training activities that provide a structure for where to go from after reading this paper. It designates stages of development, key concepts within each, and starter references. Depending on interest, not every resource needs to used, and doing further work beyond the starter literature would be critical to deepen understanding at each stage. 

We would approximately map competence in the key outcomes to the completion of the following stages in Table \ref{tbl:pathways}. A professional interested in quantum awareness would focus on Awareness and Concepts. Someone interested in being a quantum IE/OR collaborator would want to work through at least the stages up to Technical Preparation; it may be worthwhile to work through the Accessible Examples as well. To pursue full-stack competence, each stage should be completed, with specific emphasis on the target area of interest, e.g., focusing on optimization or ML. A list of general references is provided at the end.

Regarding the Technical Preparation literature, the focus is on learning and/or refreshing notation and mathematical tools that will be helpful in supporting collaborative work and understanding the subsequent resources. If you are reading selections of the \cite{Aaronson2018b} lecture notes, Lectures 1-4 (basic concepts, including notation and gates), 16-17 (computation and complexity), and 25-26 (Hamiltonians and adiabatic algorithms) would be helpful as an early reference. We refer you to any major linear algebra resource to refresh tensor products and Hermitian matrices, citing \cite{Strang2023} as an example. A list of the major concepts from linear algebra that would be helpful to know is presented in \cite{Nannicini2020} Section 2. 

Two companies noted in the software ecosystem (Section \ref{s:Hardware}), PennyLane and IBM host online tutorials focused on concepts and programming (Table \ref{tbl:pathways}: Implementation). The PennyLane Codebook Map presents a Python-based tutorial progressing from the introductory material to implementing algorithms. This resource would be a reasonable starting place to learn quantum concepts from the ground-up if you would like to skip the academic literature in earlier stages of the table and rather learn concepts via coding. The IBM Quantum Learning platform and associated Qiskit YouTube channel provide resources to on-board programmers to the Qiskit SDK and running jobs on IBM machines. A recommended starting place would be the ``Coding with Qiskit - 1.X" playlist of videos as of the time of writing (2025).

Beyond the readings and online activities, it would be worthwhile, when possible, to attend workshops, summer schools, and/or bootcamps to further technical development. Students who are interested in quantum IE/OR are encouraged to take classes in quantum information sciences. In particular, courses focusing on quantum computation, including algorithms, are likely be most relevant. Its sister subdiscipline of quantum information is often grouped with quantum computation, but on its own, it is one step removed from IE/OR work.

\begin{singlespace}
\begin{table}[ht]
\fontsize{9pt}{9pt}\selectfont
\centering
\begin{tabular}{llllll}
\hline
\textbf{Stage} & \textbf{Concept} & \textbf{References} \\
\hline
Awareness & Hardware context & \cite{Popkin2016}\\
    & Business & \cite{Ruane2022}\\
\hline
\makecell[l]{Technical\\Preparation} 
            & Linear algebra & \cite{Strang2023}\\
            & Gates &  \cite{Aaronson2018b} \\
\hline
\makecell[l]{Accessible\\Example} & \makecell[l]{Formulation,\\ Implementation,\\ and Code} & \cite{Giraldo-Quintero2022} \\
\hline
Overviews & Quantum computing & \cite{Nannicini2020} & &\\
    & Optimization and algorithms & \makecell[l]{\cite{abbas2024challenges}\\ 
    \cite{Bochkarev2025} \\ \cite{Klug2024}}  \\
    &  & \cite{bernal2020quantum} \\
    & Quantum Machine Learning & \cite{biamonte2017quantum} \\
\hline
Technical & Quantum Optimization & \cite{nannicini2024quantum}  \\
& Quantum Linear Algebra & \cite{lin2022lecture}  \\
& Ising formulations & \cite{Lucas2014} \\
            & QUBO formulations & \makecell[l]{\cite{Kochenberger2006}\\ \cite{Glover2022}}  \\
            & Grover-based optimization & \cite{Creemers2025}  \\
            & Benchmarking & \cite{Koch2025}  \\
           \hline
Implementation & General QC programming 
& \cite{IBM2025,IBM2025b}    \\
        & Specific algorithms & \cite{Abhijith2022}\\
        & QAOA Tutorial & \cite{Sturm2023}\\
\hline

\makecell[l]{General\\References} & & \cite{nielsen2010quantum}  \\
& & \cite{de2019quantum}  \\
& & \cite{mermin2007quantum} \\
\hline
\end{tabular}
\caption{Pathways.}
\label{tbl:pathways}
\end{table}
\end{singlespace}

\section{Discussion} \label{s:Disc}

Quantum has taken the world by storm, and much work remains to be done before quantum computation improves decision-making at scale. IEs, with deep expertise in bridging methodologies with applications, have the opportunity to be a valuable partner in its development. A barrier is the steep learning curve, and in this paper, we present an entryway for industrial engineers into quantum computing research. It introduces key concepts of quantum computation as well as major algorithms that may be most useful for IEs to know in the near- and medium-terms (NISQ and fault-tolerant eras, respectively). We emphasize application, a core strength of IE, and discuss the status of quantum computing hardware. Further, we present a learning pathway to enable readers to pursue research-level engagement with different aspects of quantum OR. 

For quantum computation to be useful, people working to develop its hardware, software, and use cases need to coordinate. IEs are a natural conduit between the quantum domains; the focus on methods to improve systems, decisions, and people’s lives positions IE to have an impact. In particular, the ability of IE to identify novel problems, formulate them mathematically, develop methods to solve, and then pass back to decision-makers may be a missing link to finding practical use cases where quantum computation has advantage. We will note that these skills non-trivial; many academic disciplines focus on a particular domain of real-world problems or developing methods that are abstracted from the true decisions. IEs can be the bridge.

There is a growing community of quantum IE/OR researchers. INFORMS recently launched an ad hoc committee on Quantum Computing and Operations Research \citep{Hunt2025,bernal2025what} to introduce operations researchers to quantum and to promote research. The INFORMS Journal on Computing published its first special issue focused on quantum computing in February 2025 and now has an Area for Quantum Computing. Major disciplinary conferences are hosting quantum computing tutorials and universities are hosting workshops.

The quantum computing community as a whole gathers annually at the IEEE International Conference on Quantum Computing and Engineering (aka “IEEE Quantum Week”). The major physics conference (American Physical Society [APS]) has a large quantum emphasis and then other sub-areas under quantum computation (e.g., quantum error correction and quantum information) have specialized conferences. Quantum machine learning can be found at NeurIPS (the Annual Conference on Neural Information Processing Systems) and is the focus of many specialized symposia. 

In terms of dissemination, much of the research on quantum IE/OR to date appears in physics journals. While papers are increasingly appearing in IE/OR journals, researchers may also find it worthwhile to publish in physics journals as well. These include the \textit{Physical Review} series (e.g., Parts A, C, and X); then, \textit{PRX Quantum}, \textit{NPJ Quantum}, \textit{EPJ Quantum Technology}, \textit{Physics Review Letters}, \textit{Physical Review Research}, and \textit{Quantum}. Major papers are also published in more general science journals, e.g., \textit{Nature Communications}.  
In other disciplines, \textit{IEEE Transactions on Quantum Engineering} and \textit{ACM Transactions on Quantum Computing} are also of note. Note as well that the influence of physics research has created a strong culture of preprinting quantum research papers on arXiv (generally under the quantum-physics area).

At universities, much of the quantum-related research and education was conventionally integrated within discipline-specific departments, with a handful of exceptions. Over the past few years, however, a critical mass of universities have begun to create interdisciplinary Quantum Information Science (QIS) programs, centers, and labs. These areas bring together physics, computing science, electrical engineering, and related disciplines to train students in QIS and support research on how to design quantum systems to learn about the physical world and engineer new technologies. 
As QIS develops, there is the potential for QISE (Quantum Information Science and Engineering) to bring these technologies to the public at large. 
Even if IE/OR is not a foundational area for QIS, it has the potential to be a strong partner member of these efforts, as the focus on human-centered decisions, systems, and data bridges the gap between the technology and improvement of systems in practice. 
In terms of training of IE/OR students, particularly at the graduate level, we note that the state of the quantum industry job market in three-to-five years is opaque. Many companies are currently exploring quantum computation, but medium-term job prospects will depend heavily on the on-going hardware development and realized impact on businesses. For IE/OR students aiming for industry, building expertise in quantum computation and hybrid classical-quantum systems on top of a strong foundation in classical IE/OR methodologies would position students well to navigate the potential and uncertainty in the nascent quantum computing industry.

Funding to support quantum research (at the time of writing, mid-2025) is increasing. A large portion is dedicated to hardware development, with millions needed to build a quantum computer. 
In the US, quantum is a stated priority, and federal quantum funding is distributed primarily via the National Sciences Foundation, Department of Energy, Department of Defense, and NASA. These include single and multi-investigator calls (NSF Future CoRE), as well as center-level (e.g., NSF Quantum Leap Challenge Institutes) and regional-level (NSF ENGINES) initiatives. Statewide initiatives exist as well, spanning from research, workforce, and infrastructure development.

Worldwide, public quantum funding is surging. China reported investments of \$15 billion in quantum, with billions more committed in the past two years by Japan,
the United Kingdom and Germany and the US. 
The private sector is also making massive investments. 
In the best case, when countries and countries compete, it is the advancement of human knowledge that wins. Publications transcend borders, and we learn just as much from a paper written by a research group based in Germany as we do from one based in the US. While we do not wish to understate the real impacts of where research is done (including technology transfer and workforce development), we would like to highlight the idea that “a rising tide lifts all boats.” As always, researchers should strive to conduct and report results such that it benefits the community at large, and we expect this to be especially true in quantum IE/OR.

Taken together, quantum computers are one of the next frontiers in computation. There are strong opportunities for IE/OR researchers to contribute to the forefront of quantum research - both in methodologies and applications. This paper is intended to be a starting point, and we encourage deeper study to progress from these concepts to the true advancement of knowledge. There is much work to be done, and we aim to support IE/OR researchers in amplifying their expertise in solving human-centered problems to expand the potential impact of new quantum technologies.

\if0\blind{
\section*{Acknowledgements}
We thank South Carolina Quantum for partial funding of this work.
 \fi

\section*{Supplementary Materials}

Qiskit is an open-source Python-based Software Development Kit (SDK) developed by IBM Research for writing quantum programs at various levels of abstraction \citep{Javadi-Abhari2024}. Qiskit provides tools to define quantum circuits, simulate them, and run on IBM’s quantum hardware. It includes modules for circuit optimization (the transpiler), quantum algorithms, and quantum hardware control (pulse-level programming). The design emphasizes a modular, extensible approach and is built around the circuit model of computation. For example, using Qiskit one can construct a circuit for Grover’s algorithm and execute it either on a local simulator or on an IBM Quantum cloud device.

Cirq is a Python library developed by Google Quantum AI for designing and running quantum circuits, particularly tailored to Noisy Intermediate-Scale Quantum (NISQ) devices \citep{CirqDevelopers2025}. Cirq provides primitives to write circuits and schedule them respecting hardware-specific constraints (like qubit connectivity). It is designed to interface well with Google’s quantum hardware (such as the Sycamore processor) and simulators. Cirq exposes low-level control, allowing researchers to experiment with gate sequences and error mitigation techniques on near-term devices. It has a growing collection of extensions (e.g., ReCirq for example algorithms).

PennyLane is a framework oriented toward hybrid quantum-classical machine learning and optimization developed by Xanadu \citep{Bergholm2022}.  It allows users to define quantum nodes (quantum circuits) within classical computation graphs so that quantum computations can be differentiated. It supports high-level description of quantum neural networks, quantum kernels, and variational circuits, and can interface with multiple backends (photonic simulators, qubit simulators, and real devices) via a plugin system. A central feature of PennyLane is automatic differentiation of quantum circuits, enabling gradient-based optimization of circuit parameters. This makes it a useful tool for research in quantum machine learning and variational quantum algorithms. For example, one can implement a quantum circuit in PennyLane, attach it to a PyTorch or TensorFlow optimizer, and train a quantum model similarly to training a neural network. It has an active user community in quantum machine learning.

Each of these tools lowers the barrier for industrial engineering researchers to start exploring quantum computing. They abstract away much of the physics of the hardware and provide a programmer-friendly interface. Using these libraries, researchers can construct quantum models for optimization, simulation, or machine learning problems relevant to industrial engineering, and test them on simulators or real quantum processors. The development of high-level software is critical: it enables algorithm design and testing to proceed even as hardware is still limited, and it provides a common platform for collaboration between quantum experts and domain experts in fields like operations research.

\bibliographystyle{chicago}
\spacingset{1}
\bibliography{refs}
	
\newpage
\setcounter{section}{0}
\setcounter{figure}{0}
\setcounter{table}{0}
\setcounter{page}{1}

\end{document}